\documentclass[floatfix,aps,prd,showpacs,amsmath,nofootinbib,preprintnumbers,onecolumn]{revtex4}

\usepackage{graphicx}
\usepackage{bm}

\newcommand{\figurewidthw}{3.2in}

\def\({\left(}
\def\){\right)}
\def\[{\left[}
\def\]{\right]}

\def\e{\begin{equation}}
\def\q{\end{equation}}
\def\m{\begin{eqnarray}}
\def\n{\end{eqnarray}}

\begin{document}

\title{Constraints on the cosmological parameters from BICEP2, Planck and WMAP}

\author{Cheng Cheng and Qing-Guo Huang}\email{huangqg@itp.ac.cn}
\affiliation{State Key Laboratory of Theoretical Physics, Institute of Theoretical Physics, Chinese Academy of Science, Beijing 100190, People's Republic of China}

\date{\today}

\begin{abstract}

In this paper we constrain the cosmological parameters, in particular the tilt of tensor power spectrum, by adopting Background Imaging of Cosmic Extragalactic Polarization (B2), Planck released in 2013 (P13) and Wilkinson Microwaves Anisotropy Probe 9-year Polarization (WP) data. 
We find that a blue tilted tensor power spectrum is preferred at more than $3\sigma$ confidence level if the data from B2 are assumed to be totally interpreted as the relic gravitational waves, but a scale invariant tensor power spectrum is consistent with the data once the polarized dust is taken into account. The recent Planck 353 GHz HFI dust polarization data imply that the B2 data are perfectly consistent with there being no gravitational wave signal.

\end{abstract}


\maketitle


Recently BICEP2 (B2) \cite{Ade:2014xna} detected an excess of B-mode power over the lensed-$\Lambda$CDM expectation in the range of $30< \ell<150$ multipoles. Cross correlating BICEP2 against 100 GHz maps from the BICEP1 experiment, the microwave mission by the polarized dust is disfavored at $1.7\sigma$. 
The observed B-mode power spectrum is well fitted by a lensed-$\Lambda$CDM+tensor model in which the tensor-to-scalar ratio is constrained to be 
\m
r=0.20_{-0.05}^{+0.07}, 
\label{b2r}
\n
and $r=0$ is disfavored at $7.0\sigma$. Even though there is a moderately strong tension between B2 and Planck data released in 2013 (P13) \cite{Ade:2013zuv}, the result of B2 is consistent with low-$\ell$ cosmic microwave background (CMB) data \cite{Zhao:2014rna}, including Planck TT \cite{Ade:2013zuv} and Wilkinson Microwaves Anisotropy Probe TE (WP) data \cite{Hinshaw:2012aka}. See a similar result for WMAP 7-year data in \cite{Zhao:2010ic}.

On the other hand, recently Planck team highlighted the difficulty of estimating the amount of dust polarization in the low intensity regions \cite{Ade:2014gna}. BICEP2 assumed a polarization fraction of $5\%$ for the dust which they read from the preliminary map from Planck in \cite{planck:1301}. Actually a better estimation from Planck is that the power spectrum of the dust scales as 
\m
\Delta_{\rm BB,dust,\ell}^2=\ell^2 C_{\ell}^{\rm BB,dust}/2\pi \propto \ell^{-0.3} 
\label{dust}
\n
\cite{planck:1302}. In \cite{Mortonson:2014bja,Flauger:2014qra}, the authors found that the polarized dust can also fit the data from BICEP2 quite well and $\Delta_{\rm BB,dust,100}^2\sim 0.015$ $\mu$K$^2$. Because of the absence of measurement about the foreground in the region of BICEP2, it is still hard to say what is the real origin of B-mode found by BICEP2.

If the signal from BICEP2 is confirmed to be originated from the primordial gravitational waves by the upcoming data sets, it would strongly suggest that inflation \cite{Guth:1980zm,Linde:1981mu,Albrecht:1982wi} really took place in the very early universe. Here it is very interesting for us to investigate the property of the spectrum of relic gravitational waves from current data sets including B2, P13 and WP. For simplicity, the amplitude $P_t$ of relic gravitational waves spectrum can be parameterized by  
\m 
P_t(k)=A_t \({k\over k_p}\)^{n_t}, 
\label{pt}
\n
where $n_t$ is the tilt and $k_p$ is the pivot scale. In this paper we set $k_p=0.004$ Mpc$^{-1}$. As was known, the simplest version of inflation model, i.e. the canonical single-field slow-roll inflation model, predicts $n_t=-r/8$. For $r=0.2$, $n_t=-0.025$ and a nearly scale-invariant spectrum of gravitational waves is predicted by the canonical single-field slow-roll inflation model. 
After B2 data was released, we constrained the tilt of relic gravitational wave spectrum by using B2 data only and found $n_t=-0.06_{-0.23}^{+0.25}$ \cite{Cheng:2014bma} which is nicely consistent with the inflation scenario. However, our result is different from others in \cite{lewis:2014d} and \cite{Gerbino:2014eqa,Wang:2014kqa} where an apparently blue tilted spectrum of primordial gravitational waves is preferred. Actually our method is different from theirs where they fixed all of the remaining parameters to be those from P13+WP best fit values for $\Lambda$CDM+tensor model. In this paper we will explore the tilt of tensor power spectrum more carefully.

\begin{figure*}[hts]
\begin{center}
\includegraphics[width=4.5in]{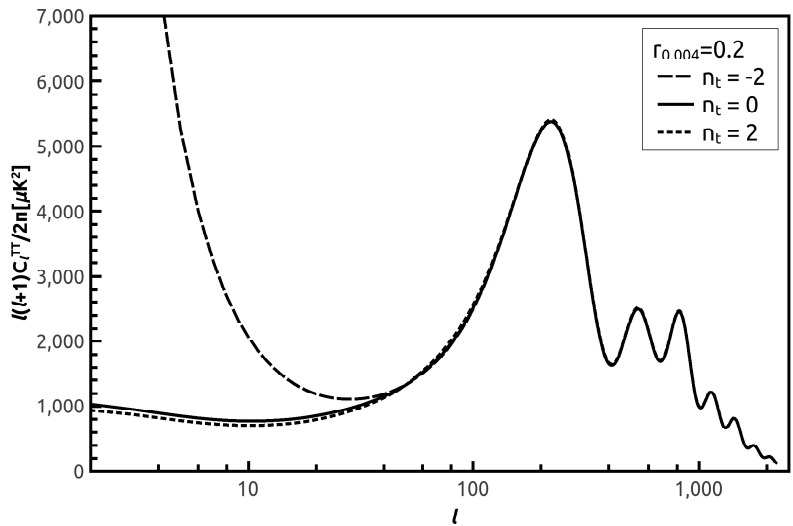}
\end{center}
\caption{The plot of $C_{\ell}^{TT}$ for different tilt of relic gravitational wave spectrum. Here $r_{0.004}=0.2$, and the black solid, dashed and dotted curves correspond to $n_t=$0, -2, +2 respectively. 
}
\label{fig:cltt}
\end{figure*}

It is well-known that the CMB power spectra generated by the primordial gravitational waves are significant only at low multipoles, e.g. $\ell \lesssim 150$. 
How the scalar perturbations and gravitational waves affect the CMB TT, TE, EE and BB spectra was illustrated in the literatures, e.g. \cite{Challinor:2004bd,Challinor:2005ye} etc. Here, for example, see Fig.~\ref{fig:cltt} where the CMB TT spectrum $C_{\ell}^{TT}$ is plotted for the tensor-to-scalar ratio at $k_p=0.004$ Mpc$^{-1}$ fixed to be $r_{0.004}=0.2$, and the black solid, dashed and dotted curves correspond to $n_t=$0, -2, +2 respectively. 
From Fig.~\ref{fig:cltt}, we see that a strongly red-tilted spectrum of primordial gravitational waves can contribute significantly to the CMB TT spectrum at very large scales (e.g. $\ell \lesssim 20\sim 30$), and then the CMB TT spectrum can provide a constraint on the negative part of $n_t$. Since the primordial gravitational waves are almost invisible on the small scales, roughly speaking, the positive part of $n_t$ is expected to be constrained by B2 data only.

First of all, by interpreting the signal from B2 totally as the primordial gravitational waves, we can constrain the cosmological parameters in the $\Lambda$CDM+r+$n_t$ model by combining B2 with P13 and WP. There are eight cosmological parameters in the $\Lambda$CDM+r+$n_t$ model: baryon density today $\Omega_b h^2$, cold dark matter density today $\Omega_c h^2$, optical depth due to reionization $\tau$, angular scale of the sound horizon at last-scattering $\theta_{\rm MC}$, amplitude of scalar power spectrum $A_s$, spectral index of scalar power spectrum $n_s$, tensor-to-scalar ratio $r$ and the tilt of tensor power spectrum $n_t$. We run CosmoMC \cite{cosmomc} and work out the constraints on the cosmological parameters in Table \ref{tab:xrnt}. 
\begin{table}[htbp]
\centering
\renewcommand{\arraystretch}{1.5}
\scriptsize 

{
 
\

\begin{tabular}{c|c}
\hline\hline
$\Lambda$CDM+r+$n_t$ & \multicolumn{1}{|c}{B2+P13+WP} \\
\hline
parameters&$68\%$ limits  \\
\hline
$\Omega_b h^2$ & $0.02200_{-0.00034}^{+0.00033}$ \\
$\Omega_c h^2$ & $0.1209\pm 0.0028$ \\
100$\theta_{\rm MC}$ & $1.04118_{-0.00064}^{+0.00065}$ \\
$\tau$ & $0.104_{-0.018}^{+0.015}$ \\
$\ln(10^{10}A_s)$ &  $3.172_{-0.032}^{+0.039}$ \\
$n_s$ & $1.035_{-0.046}^{+0.029}$ \\
\hline
\end{tabular}
}
\vspace{5mm}
{
 
\

\begin{tabular}{c|c|c|c|c}
\hline
parameters & best fit & $68\%$ limits  &$95\%$ limits & $99.7\%$ limits \\
\hline
$r$& 0.055 &[0.013, 0.072]  &[0, 0.118] & [0, 0.156] \\
$n_t$& 1.435 &[0.935, 1.933]  &[0.482, 2.389] & [0.192, 2.663] \\
\hline
\end{tabular}
}
\caption{Constraints on the cosmological parameters in the $\Lambda$CDM+r+$n_t$ model from the combination of B2+P13+WP.  }
\label{tab:xrnt}
\end{table}
From Table \ref{tab:xrnt}, we see that $n_t>0$ is preferred at more than $3\sigma$ confidence level. If it is true, the inflation model will be challenged. 

\begin{figure}[hts]
\begin{center}
\includegraphics[width=\figurewidthw]{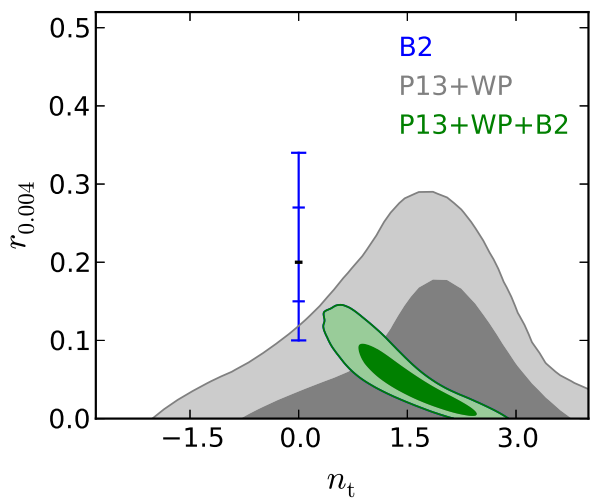}
\end{center}
\caption{The contour plot of $r$ and $n_t$ constrained by different CMB datasets in the $\Lambda$CDM+r+$n_t$ model. 
}
\label{fig:tension}
\end{figure}

Here we propose another interpretation on the above fitting results. In \cite{Ade:2014xna} BICEP2 team assumed $n_t=0$ and the tensor-to-scalar ratio is constrained to be that in Eq.~(\ref{b2r}). See the blue error bar pathology in Fig.~\ref{fig:tension}. We also notice that Planck group assumed the consistency relation for the canonical single-field slow-roll inflation model, i.e. $n_t=-r/8$, and found $r<0.11$ at $2\sigma$ confidence level. Here we relax the assumption adopted by Planck team and constrain the cosmological parameters from P13+WP in the $\Lambda$CDM+r+$n_t$ model in which $n_t$ is taken as a free parameter. The joined constraint on $r$ and $n_t$ from P13+WP is illustrated by  the gray region in Fig.~\ref{fig:tension}. From Fig.~\ref{fig:tension}, we see that the CMB TT spectrum of Planck and TE spectrum of WMAP can significantly constrain the tensor-to-scalar ratio $r$ in the region of $n_t<0$, but the constraint becomes quite loose in the region of $n_t>0$ and a broad region for $n_t>0$ is still allowed by P13+WP. It is consistent with our previous argument. From this figure, we also find that there is a strong tension between B2 \cite{Ade:2014xna} and P13 \cite{Ade:2013zuv} in the $\Lambda$CDM+r+$n_t$ model. Ignoring this tension, we can combine B2 with P13 and WP to constrain the cosmological parameters. The joined constraint on $r$ and $n_t$ from B2+P13+WP is showed by the green contours in Fig.~\ref{fig:tension}, and a blue tilted power spectrum of relic gravitational waves is preferred.  Technically, from the viewpoint of $\chi^2$ statistics, the tension between B2 and P13+WP can significantly increase the $\chi^2$ in the region of $n_t<0$, and then the fitting parameter space moves to the region of $n_t>0$. 
That is why the combination of B2 and P13+WP drives the fitting parameter space to the region of $n_t>0$. In this sense we believe that the apparently blue tilted spectrum of primordial gravitational waves in \cite{lewis:2014d,Gerbino:2014eqa,Wang:2014kqa} might not be a really physical result because it can be explained by the tension between B2 and P13+WP which should not be combined together. 

\begin{figure}[hts]
\begin{center}
\includegraphics[width=\figurewidthw]{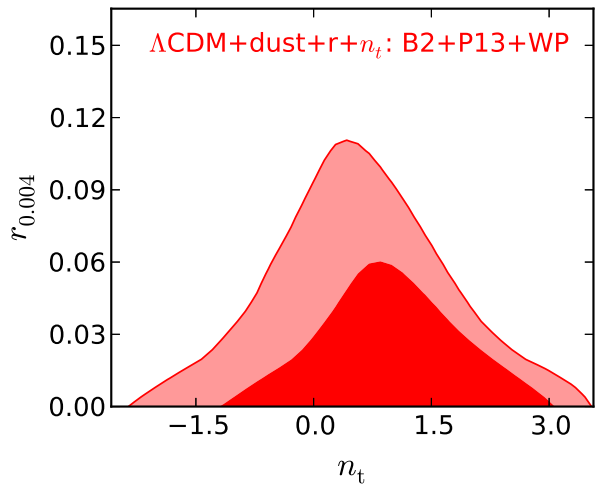}
\end{center}
\caption{The contour plot of $r$ and $n_t$ constrained by B2+P13+WP in the $\Lambda$CDM+dust+r+$n_t$ model.
}
\label{fig:drnt}
\end{figure}

In the following let's switch to the $\Lambda$CDM+dust+r+$n_t$ model. Here the contribution to $C_\ell^{\rm BB}$ from the polarized dust is estimated by Eq.~(\ref{dust}). Assuming lensing, primordial gravitational waves and dust polarization as the only factors contributing to the BICEP2 B-mode data, we constrain all of the parameters in the $\Lambda$CDM+dust+r+$n_t$ model. Our results are summarized in Table \ref{tab:drnt}. 
\begin{table}[htbp]
\centering
\renewcommand{\arraystretch}{1.5}
\scriptsize 

{
 
\

\begin{tabular}{c|c}
\hline\hline
$\Lambda$CDM+dust+r+$n_t$ & \multicolumn{1}{|c}{B2+P13+WP} \\
\hline
parameters&$68\%$ limits  \\
\hline
$\Omega_b h^2$ & $0.02208_{-0.00027}^{+0.00026}$ \\
$\Omega_c h^2$ & $0.1183_{-0.0020}^{+0.0021}$ \\
100$\theta_{\rm MC}$ & $1.04122_{-0.00059}^{+0.00061}$ \\
$\tau$ & $0.088_{-0.014}^{+0.013}$ \\
$\ln(10^{10}A_s)$ &  $3.179\pm 0.023$ \\
$n_s$ & $0.9611_{-0.0066}^{+0.0065}$ \\
$100\Delta_{\rm BB, dust, 100}^2$ ($\mu$K$^2$) &  $0.75_{-0.35}^{+0.37}$\\
$r$ & [0, 0.034] \\ 
$n_t$ & $0.841_{-0.913}^{+1.134}$ \\
\hline
\end{tabular}
}
\caption{Constraints on the cosmological parameters in the $\Lambda$CDM+dust+r+$n_t$ model from the combination of B2+P13+WP.  }
\label{tab:drnt}
\end{table}
From Table \ref{tab:drnt}, we see that there is no evidence for primordial gravitational waves and the B2 data are better interpreted as dust. \footnote{Recently extrapolating Planck HFI 353 GHz dust polarization data to 150 GHz, Planck collaboration found that the dust power is roughly the same magnitude as BICEP2 signal \cite{Adam:2014bub}, and the upper bound of the tensor-to-scalar ratio reads $r<0.083$ at $95\%$ C.L. without any evidence for the primordial gravitational waves \cite{Cheng:2014pxa}. Note that the amplitude of dust power in Table \ref{tab:drnt} is consistent with that in \cite{Adam:2014bub}. }
In this sense there is no longer tension between B2 and P13+WP and the tilt of tensor power spectrum is consistent with a scale-invariant spectrum within $1\sigma$ confidence level. See the jointed constraint on $r$ and $n_t$ in the $\Lambda$CDM+dust+r+$n_t$ model in Fig.~\ref{fig:drnt}. 

To summarize, we constrain the cosmological parameters, in particular the tilt of tensor power spectrum, in different models by combining B2 with P13 and WP. We found that a scale invariant power spectrum of relic gravitational waves is consistent with the data in the $\Lambda$CDM+dust+r+$n_t$ model where the B mode detected by BICEP2 is mainly interpreted as the polarized dust, while an apparent blue tilted tensor power spectrum is preferred in the $\Lambda$CDM+r+$n_t$ model in which the B mode is assumed to be contributed by the relic gravitational waves only. We also suggest that the apparently blue tilted tensor power spectrum can be explained by the tension between B2 and P13+WP and it might not be a real physical result.

Actually it is dangerous to use the CMB TT spectrum to constrain the properties of relic gravitational waves, e.g. the tilt $n_t$, because the relic gravitational waves only make a small contribution to the CMB temperature spectrum which can be affected by a lot of complicated physical factors, such as the baryon density today, the cold dark matter density today, the spectral index, the running of spectral index (nrun), the total mass of active neutrinos ($\sum m_\nu$), the number of relativistic species ($N_{\rm eff}$), the gravitational lensing, the abundance of light elements and so on. 
All of these complicated factors can bring strong bias on the data analysis. In a word, the polarization data like those released by B2 is still considered to be the best for us to constrain the tilt of primordial gravitational wave spectrum. 

Finally, the canonical single-field slow-roll inflation predicts a consistency relation $n_t=-r/8$. We hope that this consistency relation can be explicitly tested in the future \cite{Caligiuri:2014sla,Dodelson:2014exa}, and then the inflation model will be definitely proved.


\vspace{5mm}
\noindent {\bf Acknowledgments. }
We acknowledge the use of Planck Legacy Archive, ITP and Lenovo
Shenteng 7000 supercomputer in the Supercomputing Center of CAS
for providing computing resources. This work is supported by the project of Knowledge Innovation Program of Chinese Academy of Science and grants from NSFC (grant NO. 10821504, 11322545 and 11335012).



\end{document}